\documentclass[12pt,twocolumn,twoside,lineno]{pnas-new}

\newcommand*{\circled}[1]{\lower.7ex\hbox{\tikz\draw (0pt, 0pt)%
    circle (.5em) node {\makebox[0em][c]{\small #1}};}}
\newcommand{\ageq}{\mbox{\
\raisebox{-.9ex}{$\stackrel{\textstyle >}{\sim}$}\ }}

\newcommand{\Ma}{\mathit{Ma}}

\newcommand{\Sc}{\mathit{Sc}}

\templatetype{pnasmathematics} 

\title{Periodic bouncing of a plasmonic bubble in a binary liquid by competing solutal and thermal Marangoni forces}

\author[a,b,c,d]{Binglin Zeng}
\author[a]{Kai Leong Chong}
\author[b,d,1]{Yuliang Wang}
\author[a,e]{Christian Diddens}
\author[a,b,c,d]{Xiaolai Li}
\author[a,c]{Marvin Detert}
\author[c]{Harold J. W. Zandvliet}
\author[a,f,1]{Detlef Lohse}

\affil[a]{Physics of Fluids Group and Max Planck Center for Complex Fluid Dynamics, MESA+ Institute and J. M. Burgers Centre for Fluid Dynamics, University of Twente, P.O. Box 217, 7500AE Enschede, The Netherlands}
\affil[b]{School of Mechanical Engineering and Automation, Beihang University, 37 Xueyuan Rd, Haidian District, Beijing, China}
\affil[c]{Physics of Interfaces and Nanomaterials, MESA+ Institute, University of Twente, P.O. Box 217, 7500 AE Enschede, The Netherlands}
\affil[d]{Beijing Advanced Innovation Center for Biomedical Engineering, Beihang University, 37 Xueyuan Rd, Haidian District, Beijing, China}
\affil[e]{Department of Mechanical Engineering, Eindhoven University of Technology, P.O. Box 513, 5600 MB Eindhoven, Netherlands}
\affil[f]{Max Planck Institute for Dynamics and Self-Organization, Am Fassberg 17, 37077 G\"ottingen, Germany}

\authorcontributions{}
\authordeclaration{}
\correspondingauthor{}

\keywords{plasmonic bubbles $|$ bubble bouncing $|$ Marangoni force $|$ phase transition $|$ binary liquids}

\begin{abstract}
The physicochemical hydrodynamics of bubbles and droplets out of equilibrium, in particular with
 phase transitions, displays surprisingly rich and often counterintuitive phenomena. Here we experimentally and theoretically study the nucleation and early evolution of plasmonic bubbles in a binary liquid consisting of water and ethanol. Remarkably, the submillimeter plasmonic bubble is found to be periodically attracted to and repelled from the nanoparticle-decorated substrate, with frequencies of around a few  kHz. We identify the competition between solutal and thermal Marangoni forces as origin of the periodic bouncing. The former arises due to the selective
 vaporization   of ethanol at the substrate's
side of the bubble, leading to a solutal Marangoni flow towards the hot substrate, which pushes the bubble away. The latter arises due to the temperature gradient across the bubble, leading to a thermal Marangoni flow away from the substrate which sucks the bubble towards it. We study the dependence of the frequency of the bouncing phenomenon from the control parameters of the system, namely the ethanol fraction and the laser power for the plasmonic heating. Our findings can be generalized to boiling and electrolytically or catalytically generated bubbles in multicomponent liquids.
\end{abstract}

\begin{document}

\maketitle
\thispagestyle{firststyle}
\ifthenelse{\boolean{shortarticle}}{\ifthenelse{\boolean{singlecolumn}}{\abscontentformatted}{\abscontent}}{}

Bubbles and bubble nucleation are ubiquitous in nature and technology, e.g.\ in boiling, electrolysis, and catalysis, where the phenomena connected with them have tremendous relevance for energy conversion, or in flotation, sonochemistry, cavitation, ultrasonic cleaning, and biomedical applications of ultrasound and bubbles. This  also includes plasmonic bubbles, i.e., bubbles nucleating at liquid-immersed metal nanoparticles under laser irradiation, due to which an enormous amount of heat is produced thanks to a surface plasmon resonance \cite{baffou2013,baffou2014,lombard2015,wang2018,fang2013}. For an overview on the fundamentals of bubbles and their applications we refer to our recent review article \cite{lohse2018}. In general, in these applications the bubble nucleation does not occur in a pure liquid, but in multicomponent liquids. Because of that, various additional forces and effects come into play \cite{cates2018}, which are not relevant in pure liquids. Examples are the Soret effect \cite{bergeon1998,liu1996,maeda2011} or body forces arising due to density gradients. Once the multicomponent systems have interfaces, solutal Marangoni forces \cite{scriven1960} become relevant. The phenomena become even richer once phase transitions occur in such systems, e.g.\ solidification \cite{dedovets2018,deville2017}, evaporation \cite{christy2011,cira2015,diddens2017,diddens2017jcis,kim2018,li2018-yaxing,edwards2018,li2019prl-yaxing} or dissolution of a multicomponent droplets \cite{chu2016,dietrich2017,maheshwari2017,escobar2020}, or nucleation of a new phase such as in the so-called ouzo effect \cite{vitale2003,tan2016} or in boiling \cite{dhir1998}, electrolysis \cite{bashkatov2019,yang2018marangoni}, or catalysis \cite{manjare2012,lv2017}. Similarly, also chemical reactions occurring at the interface in a multicomponent liquid lead to spectacular effects, such as swimming droplets \cite{lauga2009,maass2016}, phoretic self-propulsion \cite{izri2014,michelin2014,moran2017,golestanian2005} or pattern formation in electroconvection \cite{mani2020}. The whole field could be summarized as physicochemical hydrodynamics, and though this is a classical subject \cite{levich1962}, it got increasing attention in recent years due to its relevance for various applications, due to new experimental and numerical possibilities, and due to the beauty of the often surprising and counterintuitive phenomena. For recent reviews on physicochemical hydrodynamics we refer to references \cite{lohse2020,dewit2020}.

To exactly analyse the various competing forces playing a role in physicochemical hydrodynamical systems, one has to strive to have simple and clean geometries, allowing for precise measurements and a theoretical and numerical approach. For example,
in refs.\ \cite{li2019prl-yanshen,li2021-yanshen} we analysed the competition between solutal Marangoni forces, gravity, and thermal diffusion by studying an oil droplet in a stably stratified liquid consisting of ethanol and water, imposing  density and surface tension gradients on the droplet. Depending on the control parameters, the droplet was either stably levitating or jumping up and down, with a very low frequency of $\sim 0.02$ Hz. Similar droplet and bubble oscillations originating from the competition between solutal Marangoni forces and gravity were observed in ref.\ \cite{schwarzenberger2015}.

In this paper, we will report and analyse another controlled physicochemical hydrodynamic bouncing phenomenon, even involving phase transitions, namely that of a nucleating plasmonic bubble \cite{fang2013,baffou2014,wang2018}, but now in an initially homogeneous {\it binary} liquid, for which the delay of bubble nucleation after turning on the laser depends on the composition of the binary liquid and the amount of dissolved gas \cite{detert2020}, (next, of course, to the power of the employed laser). As in ref.\ \cite{li2019prl-yanshen,li2021-yanshen}, we will again see a bouncing behavior, but this time on a much faster timescale, corresponding to frequencies of $\sim 10^3$ Hz. We will use this controlled physicochemical hydrodynamic system out of equilibrium to probe the competition between solutal and thermal Marangoni forces. That,
 in the presence of concentration gradients, the latter can compete with the former ones,
  is only possible thanks to the very high temperature gradients in the system of a nucleating plasmonic bubble. Under more standard conditions, such as for the evaporation of a binary droplet,
   the solutal Marangoni forces tend to be much stronger than the thermal ones \cite{diddens2017}.

We note that plasmonic bubbles are in itself very interesting with potential applications in biomedical
 diagnosis and therapy, micro- and nanomanipulation, and catalysis \cite{lapotko2009,emelianov2009,baffou2013,liu2014}. Also note that plasmonic bubbles directly after nucleation are pure vapor bubbles \cite{prosperetti2017} originating from evaporation of the surrounding liquid, but during their expansion they are invaded by dissolved gas from the surrounding liquid \cite{wang2017,li2019plasmonic,zaytsev2020,detert2020}, which in the long term crucially determines their dynamics and lifetime.

The key idea of this study here will build on the selective heating of the
liquid surrounding the plasmonic bubble, namely on the side of the plasmonic nanoparticles. This leads to very strong temperature gradients across the bubble and thus to thermal Marangoni forces and at the same time to strong concentration gradients, as the evaporation of the surrounding binary liquid is selective, favoring the liquid with the lower boiling point. Thus also solutal Marangoni forces along the bubble-liquid interface emerge. As we will see, which of these two different Marangoni forces is stronger depends on time and bubble position, leading to an oscillatory or bouncing bubble behavior.

\section*{Experimental setup and findings}
The experimental setup is as follows: The gold nanoparticle decorated substrate was put in a quartz glass cuvette (10$\times$10$\times$45 mm) filled with an ethanol-water binary liquid of varying composition (see Fig. S1 in Supplementary Material for more details on sample preparation). A continuous laser (Cobolt Samba) of 532 nm wavelength was used for substrate irradiation from the bottom side. The laser power was controlled by using a half-wave plate and a polarizer and measured by a photodiode power sensor (S130C, ThorLabs). Two high speed cameras were installed in the setup to monitor the dynamics of the generated bubbles. One (Photron SA7) was equipped with a 5$\times$ long working distance objective (LMPLFLN, Olympus) for bottom view imaging, and the other one (Photron SAZ) with a 10$\times$ long working distance objectives and operated at 200 kfps for fast imaging (see Fig. S2 in Supplementary Material for the sketch of the setup).

\begin{figure}[h!]
 \centering
\includegraphics[width=0.5\textwidth]{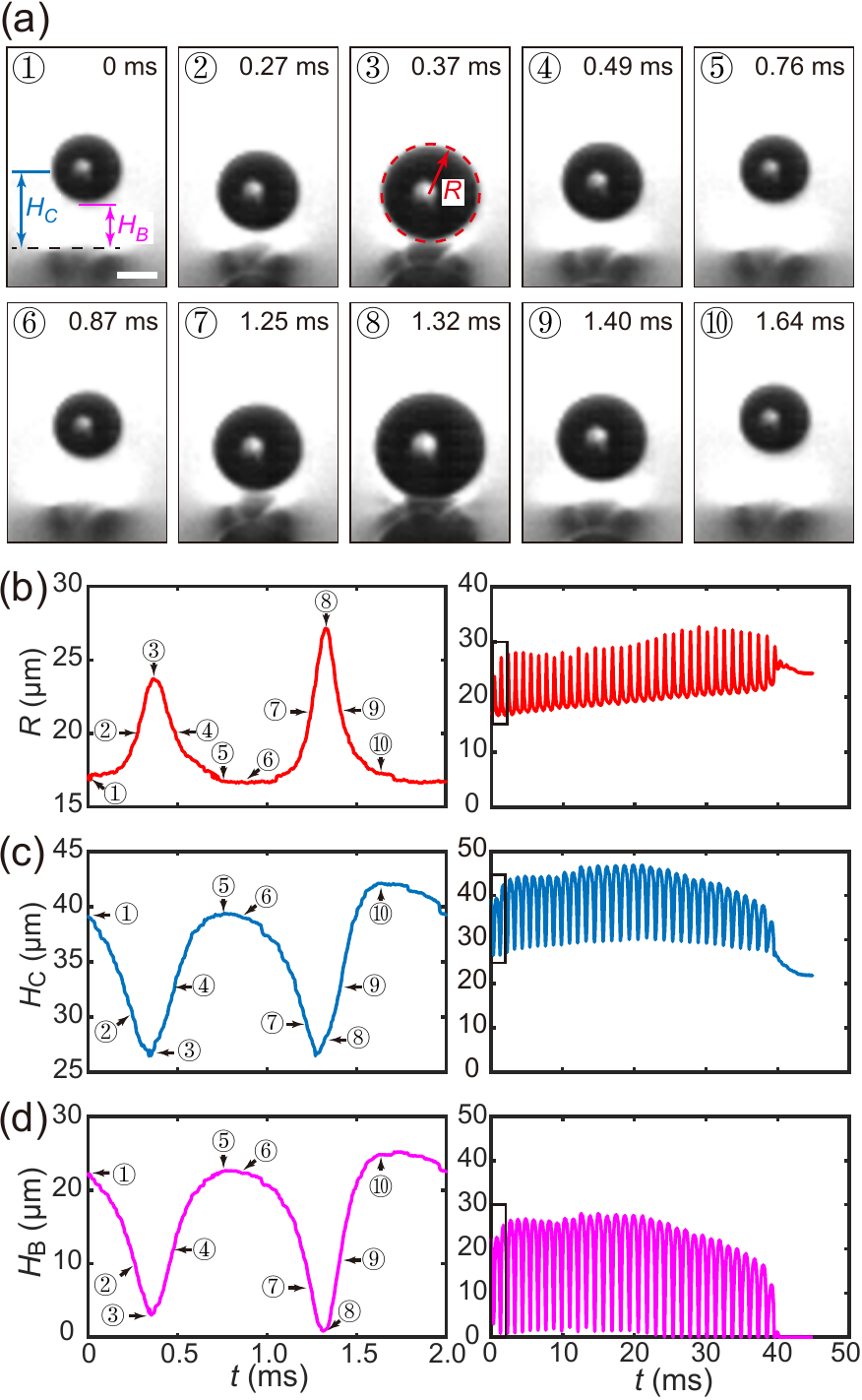}
\caption{Successive bouncing of a plasmonic bubble in an ethanol-water binary liquid with ethanol concentration at
$n_e = 75$  wt\% and laser power $P_\ell = 50$ mW under continuous laser irradiation.
(a) Experimental snapshots of the bubble bouncing for the first two cycles. The scale bar is 20 $\mu$m. The instantaneous bubble radius $R$, the center position $H_{C}$, and distance $H_{B}$ from the bottom
of the bubble to the substrate with plasmonic particles are defined in the 3rd and 1st image, respectively. Obviously, $H_C = H_B + R$.
(b-d) Left: The bubble radius $R$, the center position $H_{C}$, and the distance $H_{B}$ from the bottom versus time $t$ for the first two cycles of the bouncing process. Right: The same on a much longer time scale. The area selected by the rectangular boxes is the time interval shown in the left figures. This particular bubble bounces 34 times within 40 ms, corresponding to a mean bouncing frequency of $f_b = 0.85$kHz .
 }
\label{fig1}
\end{figure}

A series of typical snapshots of the bubble motion in an ethanol-water binary liquid (ethanol: 75 wt$\%$) for two early bouncing cycles (about 19 ms after nucleation of the plasmonic bubble) is shown in Fig.\ref{fig1}(a). The radius $R(t)$, the distance of the bubble center from the plasmonic-particle-decorated substrate $H_C(t)$, and the distance between the bubble bottom and the substrate $H_B(t)$
 are plotted in Figs.\ \ref{fig1}(b)-\ref{fig1}(d), with two different temporal magnifications.
In the first three frames of Fig.\ \ref{fig1}(a), $\circled{1}$ - $\circled{3}$, the bubble is moving towards the substrate and expanding under laser irradiation. At 370 $\mu$s, the bubble bottom touches the substrate and the bubble radius reaches a maximum value of 24 $\mu$m. Subsequently, the bubble jumps up and at the same time shrinks ($\circled{3}$ - $\circled{5}$). At 870 $\mu$s, the bubble radius has its minimum of 17 $\mu$m and the distance between its bottom and substrate reaches 22 $\mu$m ($\circled{6}$). After that, it moves again towards the substrate for another cycle ($\circled{6}$ - $\circled{10}$). In the entire process, the bubble continuously bounces towards and away from the substrate for 34 times. Finally, it remains sitting on the substrate and enters its stable and monotonous growth phase.
For details of the whole bouncing process, please refer to Movie S1 in the Supplementary Material.

 \begin{figure*}[h!]
 \centering
 \includegraphics[width=0.85\textwidth]{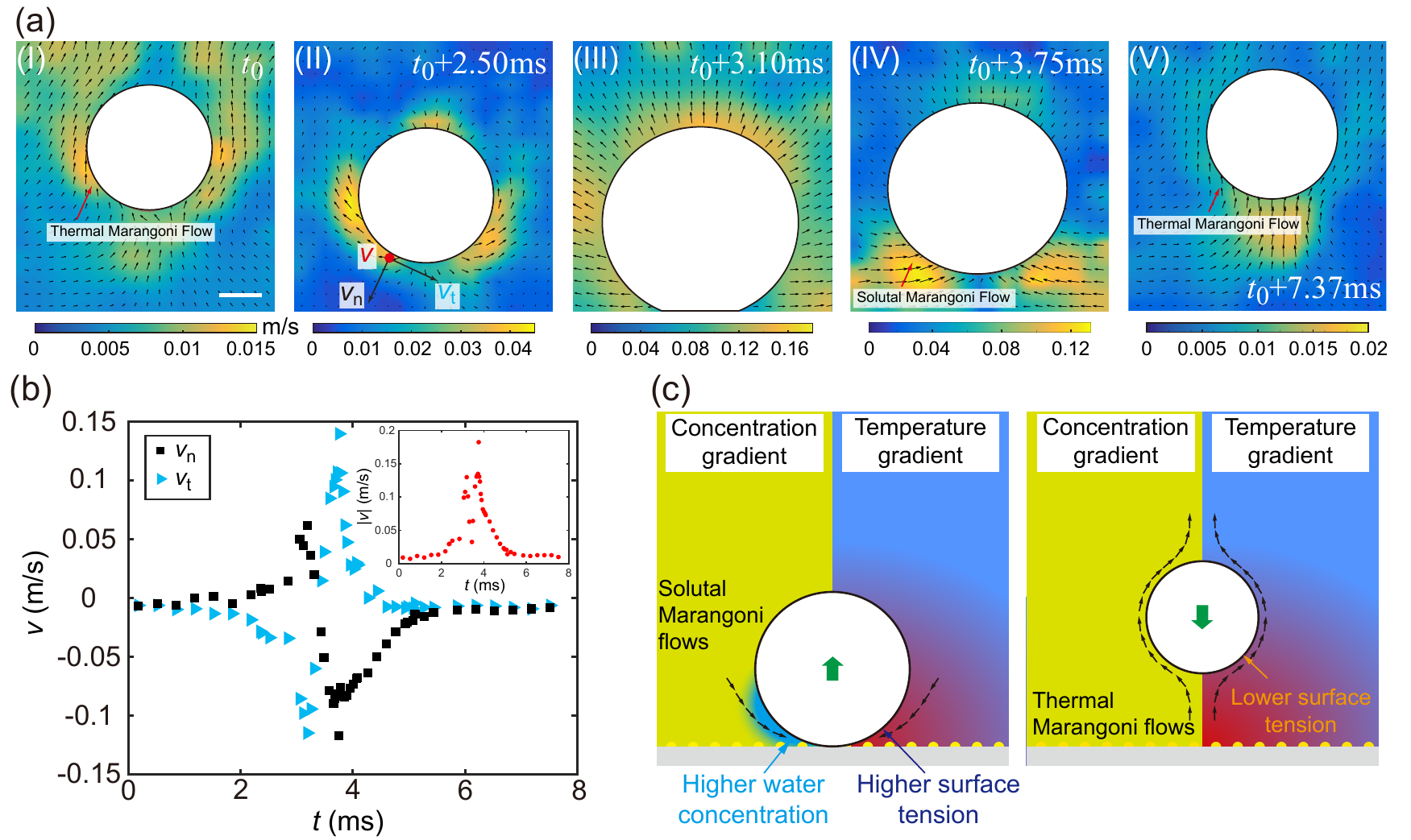}
 \caption{
 (a) PIV measurements of the plasmonic bubble for one bouncing cycle. The scale bar is 50 $\mu$m. In the experiment, the control parameters (initial concentrations in the binary mixture) are the same as in Fig. \ref{fig1}.
 (I) The temperature gradient generates thermal Marangoni flow, which drives the bubble downwards.
 (II) Shortly after the bubble starts to move downwards, the thermal induced Marangoni flow becomes stronger.
 (III) The bubble hits the superheated surface, leading
 to  vaporization of (mainly) ethanol
 near the substrate into the bubble  and corresponding bubble growth.
The bubble expanding flow can  be
 recognized from the  velocity vectors pointing radially  outwards from the bubble's center.
 (IV) The selective
  vaporization of ethanol from
   the binary liquid leads to
    a water rich area near the substrate, inducing a
     solutal Marangoni flows towards the  bubble's bottom and correspondingly  an upwards   push of the bubble.
 (V) The Marangoni flow is dominated by the
  temperature gradient again when the  bubble no longer feels a concentration gradient.
  In this cooler region it shrinks again. Note that there is a sideways motion of the bubble in subfigure V owing to the asymmetry in the solutal Marangoni flow (see subfigure IV).
 (b) Measured velocities as a function of time $t$ from the PIV measurement shown in (a). The given values refer to the positions shown in the subfigure in the lower right corner of Fig. \ref{fig2}a-II: $V_n$ (black) and $V_t$ (blue) are the normal and the tangential velocity component, respectively. The subfigure in the upper right corner displays the velocity magnitude $|V| = \sqrt{V_n^2 + V_t^2}$ as
  function of time $t$.
(c) Schematic diagrams of the bubble bouncing mechanism. The left figure sketches
 the repelling phase when the solutal Marangoni force dominates and the right one
 the attracting phase when the thermal Marangoni flow dominates.
 }
 \label{fig2}
 \end{figure*}

To understand the origin of the bouncing behavior, we investigate the flow dynamics by particle image velocimetry (PIV). Fluorescent particles with a diameter of 1.1 $\mu$m and a concentration of 100 $\mu$g/mL were added to the ethanol-water binary liquid, and illuminated with a laser (see Supplemental Material for more details and results). The results of the PIV measurements for one jumping cycle are shown in Fig.\ \ref{fig2}a.
In Fig. \ref{fig2}b we also show
 the measured time dependence of the tangential and the normal velocity components on the surface of the bubble on the side towards the heated substrate (see sketch in Fig.\ \ref{fig2}a-II), see also
 Fig. S3 and Movie S2 in the Supplementary Material. Flow illustrations during the repelling and attracting phase are shown in the left and right panel of Fig. \ref{fig2}c, respectively.

\section*{Interpretation of the findings}
Our interpretation of the measurements is as follows: we start the cycle at moment $\circled{3}$ (Fig. \ref{fig1}a), when
 the bubble has its maximal extension
  and is (nearly) in touch
  (the minimum distance between the bubble and the wall is in range from 0 to 3 $\mu$m, depending on the values of the control parameters)
  with the plasmonic-particle-decorated and thus hot substrate. Correspondingly, on the bubble side towards the hot substrate there is selective evaporation of ethanol, due to the lower boiling point of ethanol as compared to water (Fig. S4 in Supplementary Material). This leads to a higher relative water concentration at that side of the bubble and thus to a solutal Marangoni flow along the bubble interface towards the substrate (Fig.\ \ref{fig2}c left), see Fig.\ \ref{fig2}a.
This flow pushes the bubble up, away from the substrate. As there the temperature is lower, the
vapor inside the bubble partially condenses,
preferentially at the side away from the substrate,
and the bubble shrinks and experiences a decreasing concentration gradient along its interface and thus a weaker Marangoni force.
At moment $\circled{6}$ the plasmonic bubble has reached its minimal size. However, it still experiences a strong temperature gradient as the thermal diffusivity is much faster than the solutal diffusivity.
The ratio  of the two diffusivities is the Lewis number $Le$, which
 is about 110  for ethanol at the temperatures we consider here. As a result, the thermal boundary layer is much thicker than the concentration boundary layer. The strong temperature gradient across the bubble with higher temperature (and thus lower surface tension) on the bubble side towards the substrate and lower temperature and thus higher surface tension on the opposite side leads to a thermal Marangoni flow along the bubble interface, away from the substrate. This flow pushes the bubble towards the substrate (moment $\circled{7}$ in Fig. \ref{fig1}a and right panel of Fig. \ref{fig2}c) where it arrives at moment $\circled{8}$, which is equivalent to the starting moment $\circled{3}$ of the cycle.

\begin{figure}[h!]
 \centering
 \includegraphics[width=0.5\textwidth]{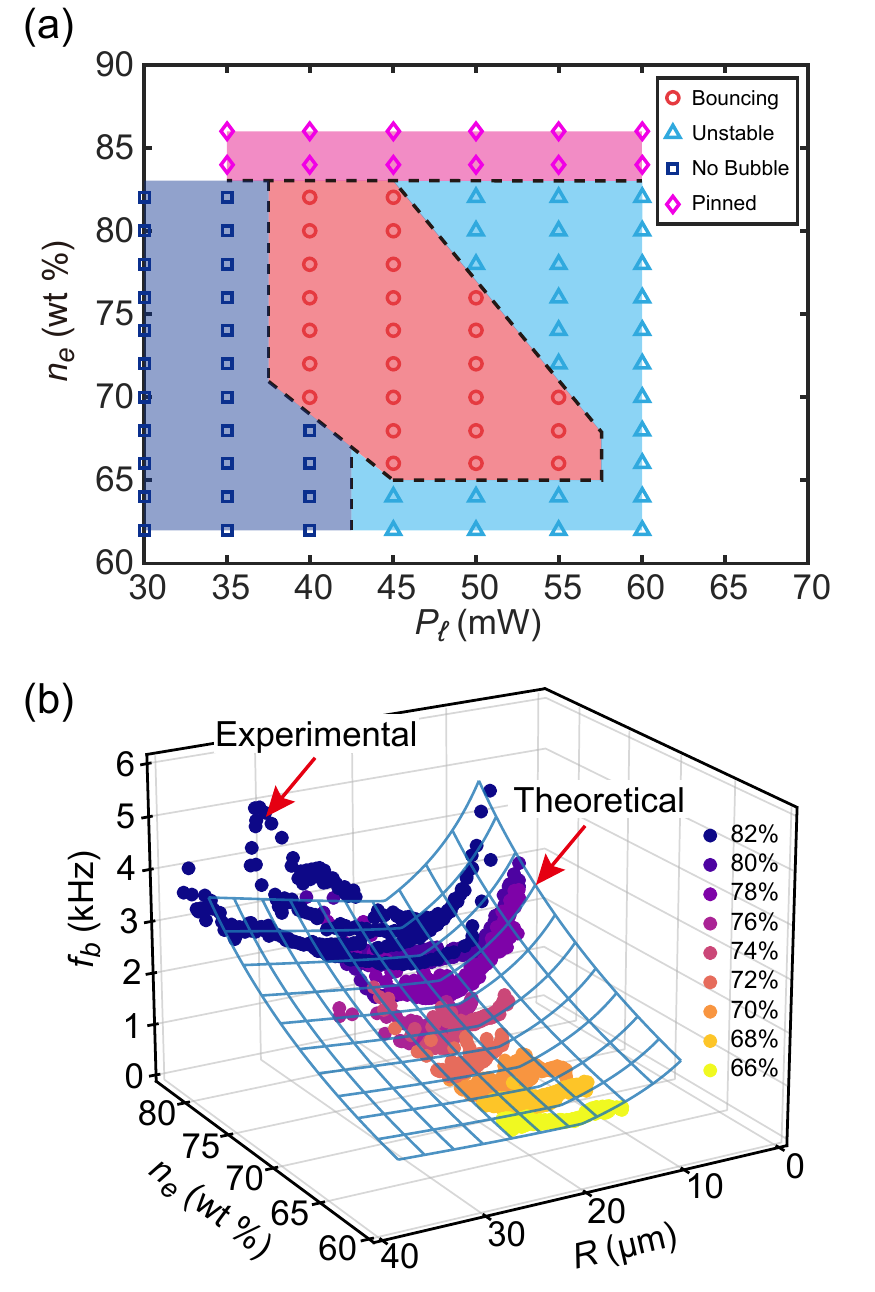}
 \caption{
 (a) Phase diagram for the
 bubble behaviors at the gold nanoparticle decorated sample surface upon continuous wave laser irradiation. The control parameters are
 the  ethanol concentration $n_e$ and the laser power $P_\ell$.
 Bubble bouncing takes place in the red regime where  40 mW $< P_\ell <$ 55 mW and 66 wt \% $< n_e <$ 82 wt \%.
 (b) Experimentally measured (data points, $P_\ell$ = 45 mW) and theoretically estimated (transparent curved surface) bubble bouncing frequency $f_b$ as function of the ethanol concentration $n_e$ and the bubble radius $R$.
 Both exhibit similar trends. For the theoretical curve the prefactor $k=23$ of the Prandtl-Blasius-Pohlhausen relation has been
 adapted to the
 experimental results.
 }
 \label{fig3}
 \end{figure}

We note here that in the bubble bouncing process gravity does not play any role, in contrast to the bubble bouncing experiment of ref.\ \cite{li2019prl-yanshen} that is done on much larger length scales.
This can easily be seen  from calculating
 the ratio $\Delta\rho gR^2/\Delta\sigma$
 between gravity and the Marangoni force \cite{lohse2020}. For a bubble with radius $R = $ 2$\times$10$^{-5}$m, a density
 of $ \rho \approx 900 $kg/m$^3$ for the water-ethanol mixture
 and $\Delta\sigma \approx 0.03 $N/m  as approximate difference in surface tension between water and ethanol we obtain $\approx 10^{-4}$ for
 this ratio, i.e., negligible gravity.
 To further confirm that gravity does not play any role during this process, the sample substrate was placed upside down on the top window of the cuvette and a bubble was produced below the sample substrate. The same bouncing phenomenon is observed (see Movie S3 in Supplemental Material for details).

\section*{Parameter space}
To get more insight into the competition of the solutal and thermal Marangoni forces during the bouncing process, systematic experiments were conducted by tuning the laser power $P_\ell$ from 30 mW to 60 mW with ethanol concentrations $n_e$ from 62 wt\% to 86 wt\%. In the phase diagram shown in Fig. \ref{fig3}a, only in the region filled by red circles (40 mW $< P_\ell <$ 55 mW and 66 wt \% $< n_e <$ 82 wt \%) bubble bouncing takes place. Beyond this parameter range, there are three
 more phases in the phase diagram: unstable bubbles, no bubbles and pinned bubbles, respectively (see Movie S4 in Supplemental Material for comparison). For bouncing bubbles, we find that the bouncing frequency $f_b$ is strongly influenced by the ethanol concentration $n_e$ and the bubble radius $R$, as shown in Fig. \ref{fig3}b. It is found that $f_b$ increases monotonously with increasing $n_e$, while it depends non-monotonously on $R$. For $R \lesssim$ 15 $\mu$m, $f_b$ decreases with increasing $R$, while for $R \gtrsim$ 15 $\mu$m, $f_b$ increases with increasing $R$. (See Movie S5 in the Supplemental Material for comparisons with different $n_e$.)

\section*{Theoretical model}
To understand the observed dependence of $f_b$ on $n_e$ and $R$, we propose a simple theoretical model of the bouncing process. The essence of the competition between the solutal and the thermal forces acting on the bouncing plasmonic bubbles can be described by a simple effective point force ODE model for the bubble's center position $z= H_c(t)$ and its radius $R(t)$, with varying added mass of the bubble plus fluid acceleration, in the spirit of refs.\ \cite{magnaudet1998,magnaudet2000,rensen2001,toegel2006}, see also section IX of ref.\ \cite{lohse2018}.
Within such a model the normal (to the substrate) force on the moving bubble can in principle
be described by \cite{magnaudet1998}
\begin{eqnarray}
\centering
\label{force-balance}
F(t)&=& - 4\pi\rho\nu R(t){\dot z}(t)+\frac{4}{3}\pi\rho R(t)^3
\ddot z (t)
+ \frac{2}{3}\pi \rho\left( \frac{ d\left[R(t)^3{\dot z}(t) \right]}{dt}
+2R(t)^3 \ddot z(t) \right) \nonumber \\
&+&8\pi\rho\nu\int_0^t\exp\left[ 9\nu \int_\tau^t R(t')^{-2}dt' \right]
\times \mbox{erfc}\left[ \sqrt{9\nu\int_\tau^t R(t')^{-2}dt'}\right]
 \frac{d\left[ R(\tau){ \dot z}(\tau) \right]}{d\tau}d\tau \nonumber \\
 &+& F_{s-M}(R(t), z(t)) + F_{th-M} (R(t), z(t)) .
\end{eqnarray}

Here it has been assumed that the bubble dynamics is not affected by  wall effects, i.e., the prefactors
of the various force contributions refer to the case of a bubble in the bulk.

The first term is the Stokes drag force on the bubble; taking the Stokes drag
 is justified as the bubble's translational Reynolds number
$Re_t(t) = \dot H_c(t) R(t) /\nu   $ remains
smaller than 1. (In fact, an upper estimate with $\dot H_c\approx$ 0.06 m/s, $R \approx 2\times10^{-5}$ m, and
$\nu=2.2\times10^{-6} $ m$^2$/s for the kinematic viscosity of the liquid gives $Re_t = 0.54$).
The other terms in the first row reflect added mass plus liquid acceleration;  the term in second row reflects
 the Basset history force, 
 and the terms in the last row are
 the solutal and thermal Marangoni forces, respectively.
If the bubble is fully immersed in the boundary layers, the two Marangoni forces can be modeled as
$F_{s-M} \sim + \gamma_{s}R^{2}$ and $F_{th-M} \sim - \gamma_{th}R^{2}$ \cite{lohse2020, li2021-yanshen}, where $\gamma_{s}$ and $\gamma_{th}$ are the surface tension gradients caused by concentration difference and temperature difference along the vertical direction, respectively.
Both are assumed to be constant within the respective boundary layers.

The concentration boundary layer thickness
$\lambda_{s}$ is much smaller
 than the thermal boundary layer thickness
 $\lambda_{th}$. It even holds
 $\lambda_s < R$ so that the
 solutal Marangoni force is relevant only
  near the substrate. This
  also becomes
  evident from the PIV experiments, see Fig. \ref{fig2}a-IV. Therefore, we consider the solutal Marangoni force on the bubble as a short impulse,
  i.e.,  like shooting a bullet. Correspondingly, we call our simple model the ``bullet model''.
  After the  short impulse, the bubble's motion is dominated by the thermal Marangoni force
  $ \sim - \gamma_{th} R^{2}$ and
  the Stokes drag force $- 4\pi\rho\nu R \dot{z}(t)$. Accordingly, the dynamical equation for the
   bouncing  bubble can be simplified to
\begin{equation}
\frac{10}{3}\pi\rho R^3 \ddot{z}(t)=-4\pi\rho\nu R \dot{z}(t)-\gamma_{th} R^{2}.
\label{zddot}
\end{equation}
Here we took the bubble radius as constant for simplicity,
which  is reasonable  since the radius only considerably
changes within a small fraction of time ($\sim15$\%) during the  bouncing cycle, see Fig.\ S6 of the
Supplemental Material.
The prefactors of the first two terms in eq.\ (\ref{zddot}) again refer to the case of a bubble in the bulk and ignore possible
modifications of the added mass coefficient and the drag coefficient due to wall effects, whereas
for the third term we put the prefactor to 1. Both
 is justified as later an unknown prefactor has to be introduced anyhow, in which such prefactors can be absorbed, see equation
 (\ref{t-duration-dimensionless}) below.

The  initial condition for the position is  $z(0)=0$ and the initial velocity $\dot{z}(0)$
follows from equating the work done by the solutal Marangoni force  $W_{s-M} = \int F_{s-M}(z)  dz $  to  the initial kinetic energy
 $E_k = \frac{1}{2} (\frac{10}{3}\pi\rho R^3) \dot{z}(0)^{2}$ for the moving liquid around the bubble. As $\lambda_s < R$ , the solutal Marangoni
 force acts only in the concentration  boundary layer and the work it performs can be estimated as $W_{s-M} = \gamma _s \lambda_s^3$ .
With these initial conditions for $z(0)$ and $\dot{z}(0)$, eq. (\ref{zddot}) can be solved analytically to finally arrive at an
 implicit equation for  the duration time $t_{tur}$ for the bubble to return to its initial position (which is the inverse bouncing
 frequency $f_b$)
\begin{equation}
 t_{dur}= f_b^{-1} =
 \frac{5 R^{2}}{6 \nu}\left(1+\sqrt{\frac{48 \pi}{5} \frac{\gamma_{s}}{\gamma_{th}} \frac{\rho \nu^{2}}{\gamma_{th} R^{2}} \frac{\lambda_{s}^{3}}{R^{3}}}\right)\left(1-e^{-\frac{6 \nu}{5 R^{2}} t_{dur}}\right).
\label{t-duration}
\end{equation}

The  thickness $\lambda_s$ of the concentration boundary layer  is determined by the
diffusion equation for a scalar field \cite{schlichting1979}
as $\lambda_s \sim k ~ \hbox{min}(R, R_\ell)  / \sqrt{ \Sc} $, where
$Sc$  is the
Schmidt number  $Sc = \nu /D$, with $D\approx10^{-9}$ m$^2$/s  the mass diffusivity of water in ethanol.
The expression for $\lambda_s$ takes into account that,
if  the bubble becomes larger than the radius  $R_{\ell}\approx 20 \mu$m of  the heated laser spot,
the relevant length scale must be  the concentration boundary layer thickness around that laser spot.
 The prefactor $k$ is {\it a priori}   unknown
and is  adapted to the experimental data,  see figure \ref{fig3}b.
With $k= 23$ they can be well described.
 More details on the bullet model are given
in the  Supplemental Material, see in particular Section 6 and  figure S7.

It is instructive to identify the appropriate dimensionless control parameters of the bouncing plasmonic bubble.
These are the solutal and thermal Marangoni numbers
$\Ma_{s} \equiv  R^2  \gamma_{s}/(\rho \nu D)$ and $\Ma_{th} \equiv R^2  \gamma_{th}/(\rho \nu D)$, respectively,
the Schmidt number $Sc$, and the ratio $R_\ell / R$ of the
laser spot radius  and the bubble radius.
For the definition of the Marangoni numbers, we define the surface tension gradients
$\gamma_{s}$ and $\gamma_{th}$ {\it separately}
 for the concentration  and the temperature differences
 within the respective boundary layers. This can reasonably be done
 as  the solutal boundary layer thickness is much thinner than thermal one.
 With these definitions and using the viscous time scale $R^2/\nu$ and the bubble radius $R$ as reference scales,
 eqs.\ (\ref{zddot}) and (\ref{t-duration}) can be rewritten as
\begin{equation}
\frac{10}{3} \pi \tilde{z}^{\prime \prime}=-4 \pi \tilde{z}^{\prime}-\frac{\Ma_{t h}}{\Sc},
\label{force-balance-dimesionless}
\end{equation}
\begin{equation}
\tilde{t}_{dur}= \tilde f^{-1}_b =
\frac{5}{6}\left(1+\sqrt{k^3 ~  \frac{48 \pi}{5} ~ \frac{\Ma_s }{\Ma_{th}^{2}\Sc^{1/2}}
~ \hbox{min} \left(1, {R_\ell \over R}\right)
}\right)
\left(1-e^{- 6 \tilde{t}_{dur}/5}\right) ,
\label{t-duration-dimensionless}
\end{equation}
reflecting that
the bubble bouncing period indeed only depends on the four dimensionless groups
 $\Ma_{th}$, $\Ma_s$, $\Sc$,  and $R_{\ell }/R $, and that thermal and solutal Marangoni forces compete with each other.
  It also shows that there is only one free dimensionless
 parameter, $k$, in which possibly different prefactors in the terms of eq.\ (\ref{zddot}) can all be absorbed.

 The bullet  model allows us to understand the physics of the experimentally found
 parameter dependences of figure \ref{fig3}b.
First, the monotonous increase of the bouncing frequency
 $f_b$ on the ethanol concentration
  $n_e$ is due to the fact that, in the experimental parameter space of 62 wt \% $< n_e <$ 86 wt \%, a higher $n_e$ leads to a smaller surface tension gradient $\gamma_{s}$.
This results in a lower kinetic energy $E_{k}$ after the solutal Marangoni impulse, and thus to a higher bouncing frequency
$f_b$ as the bubble is pulled back earlier by the thermal Marangoni force.
Next, the nonmonotonous dependence of the bouncing frequency
$f_b$ on the bubble radius
 $R$ is due to the transition of the concentration boundary layer thickness
 $\lambda_{s}$ from being determined by the bubble radius $R$ to being determined by the laser spot radius $R_\ell$.
For a tiny bubble with $R < R_{\ell}$, an  increase of $R$ results in an increase of the
kinetic energy ($E_{k}\sim\gamma_{s}\lambda_{s}^{3}$), leading to a decrease of $f_b$. In contrast,
for $R \geq R_{\ell}$, the boundary layer thickness
$\lambda_{s}$ is limited by $R_{\ell}$ and $E_{k}\sim\gamma_{s}\lambda_{s}^{3}$ remains nearly constant.
However, the added mass of the bubble does increase with increasing $R$.
This leads to a decrease of the initial velocity $\dot z(0)$ and an increase in the bouncing frequency  $f_b$.

Finally, we put the experimentally found  phase diagram of figure \ref{fig3}a
-- showing that bubble bouncing can only take place if  both
 the laser power $P_\ell$ and the ethanol concentration $n_e$ are  in a moderate range --
into the context of our model:
For the no-bubble-regime at low laser powers $P_\ell$,
the laser power is below the  threshold
 for the nucleation of the plasmonic bubble at the given ethanol concentration \cite{wang2018}.
Too high laser powers however lead to an unstable bubble behavior (``unstable regime''):
For large ethanol concentration $n_e$,  due to the massive vaporization of ethanol, new bubbles can already
 form before the thermal Marangoni forces have pulled back the former bubble
 to the substrate. This can lead to bubble collisions and thus instability.
 On the other hand, for ethanol concentrations $n_e \lesssim $ 66 wt\%, the strong solutal Marangoni force (reflected  in a
 large $\gamma_s$)
 leads to a  very high jumping height of the bubble, so that also here
 a new bubble can form before the return of the former bubble, see equation (\ref{t-duration}).
Finally, in the  pinned regime  $n_e \ageq $ 82\%,
the ethanol concentration difference is so small that the
 solutal Marangoni force directing  away from the substrate is too weak to depin the bubble from  the substrate.

\section*{Conclusions and outlook}
In conclusion, we have experimentally and theoretically studied the competition between thermal and solutal Marangoni forces acting on a nucleating plasmonic bubble in a binary liquid. This competition leads to a periodic bouncing of the bubble towards and away from the surface
 with a frequency of several  kHz. The phenomenon not only exemplifies the richness of phenomena which can occur in the physicochemical hydrodynamics of bubbles far from equilibrium and with phase transitions, but it is also very relevant way beyond the nucleation and early life of plasmonic bubbles in binary liquids:
 It can straightforwardly be generalized to the nucleation and early life of vapor bubbles in boiling phenomena in binary liquids or to electrolytically or catalytically generated bubbles in multicomponent liquids. For applications, the bouncing of these bubbles can presumably be a considerable advantage, leading to much better mixing in the liquid close to the substrate and therefore for higher efficiency in processes like electrolysis and catalysis.

\bibliography{nanobubble-and-inkjet-literatur}

\end{document}